\documentclass[prl,twocolumn,showpacs,superscriptaddress,nofootinbib,floatfix,10pt]{revtex4-1} 
\usepackage[utf8]{inputenc}
\usepackage{color}
\usepackage{amsfonts,amsmath,amssymb}
\usepackage{graphicx}
\usepackage{bm}
\usepackage{hyperref}
\usepackage{color}
\usepackage{epstopdf}

\newcommand{\ddstaro}{D^0\bar D^{*0}}
\newcommand{\dstardstar}{D^{*0}\bar D^{*0}}
\newcommand{\X}{X(3872)}

\begin{document}

\title{Novel Method for Precisely Measuring the $X(3872)$ Mass}

\author{Feng-Kun Guo}\email[E-mail address: ]{fkguo@itp.ac.cn}

\affiliation{CAS Key Laboratory of Theoretical Physics, Institute of Theoretical Physics, Chinese Academy of Sciences, Beijing 100190, China}
\affiliation{School of Physical Sciences, University of Chinese Academy of Sciences, Beijing 100049, China
}

\begin{abstract}
The $\X$ is the first and the most interesting one amongst the abundant $XYZ$ states. Its mass coincides exactly with the $D^0\bar D^{*0}$ threshold with an uncertainty of 180~keV. Precise knowledge of its mass is crucial to understand the $\X$. However, whether it is above or below the $\ddstaro$ threshold is still unknown. We propose a completely new method to measure the $\X$ mass precisely by measuring the $\X\gamma$ line shape between 4010 and 4020~MeV, which is strongly sensitive to the $\X$ mass relative to the $\ddstaro$ threshold due to a triangle singularity. This method can be applied to experiments which produce copious $\dstardstar$ pairs, such as electron-positron, proton-antiproton and other experiments, and may lead to much more precise knowledge of the $\X$ mass.
\end{abstract}

\date{\today}

\maketitle

\vspace{2cm}

{\it Introduction.}---One of the current mysteries in the physics of the strong interaction is how the plethora of the so-called $XYZ$ states can be understood. They correspond to tens of structures in the heavy-quarkonium mass region reported in various experiments with properties beyond the standard heavy quarkonia (for reviews, we refer to Refs.~\cite{Swanson:2006st,Voloshin:2007dx,Chen:2016qju,Hosaka:2016pey,Lebed:2016hpi,Esposito:2016noz,Guo:2017jvc,Ali:2017jda,Olsen:2017bmm,Karliner:2017qhf,Yuan:2018inv,Kou:2018nap,Cerri:2018ypt}). The start of the whole story is the $\X$ discovered by the Belle Collaboration in 2003~\cite{Choi:2003ue}. Its most salient property is that the mass coincides exactly with the $\ddstaro$ threshold. Taking the ``OUR AVERAGE'' values in the 2018 version of the Review of Particle Physics~\cite{Tanabashi:2018oca} for the masses:
$m_{D^0}=(1864.84\pm0.05)$~MeV, $m_{D^{*0}}=(2006.85\pm0.05)$~MeV, and 
$m_{X}=(3871.69\pm0.17)$~MeV from the $J/\psi \pi^+\pi^-$ and $J/\psi\omega$ modes, one gets 
\begin{equation}
\delta \equiv m_{D^0}+m_{D^{*0}}-m_{\X}=(0.00\pm0.18)~\text{MeV}\,.
\end{equation}
This, together with the large branching fraction of the $\X$ into $\ddstaro$~\cite{Tanabashi:2018oca} (throughout the Letter, we use $\ddstaro$ to represent the positive $C$-parity combination of $\ddstaro$ and $D^{*0}\bar D^0$), indicates the paramount role of the $D^0\bar{D}^{*0}$ in the $\X$ dynamics. The quantum numbers of the $\X$ have been measured to be $J^{PC}=1^{++}$~\cite{Aaij:2013zoa}. The properties and calculations in the hadronic molecular scenario have been nicely reviewed in Ref.~\cite{Kalashnikova:2018vkv}. 
Precise information of $\delta$ and the width of the $\X$ are crucial in understanding the $\X$ (see, e.g., Refs.~\cite{Hanhart:2007yq,Artoisenet:2010va}). There is only an upper limit on its width as $\Gamma_{\X}<1.2$~MeV at a 90\% confidence level, and experimental efforts are going on towards measuring the width precisely~\cite{PANDA:2018zjt}.
Despite the precise measurements of the masses of the $\X$, $D^0$ and $D^{*0}$, we still do not know whether the $\X$ mass is above or below the $\ddstaro$ threshold. In this Letter, we propose a novel method for measuring the mass of the $\X$ relative to the $\ddstaro$ threshold, i.e., $\delta$, with a high precision.

\begin{figure}[b]
   \includegraphics[width=0.45\textwidth]{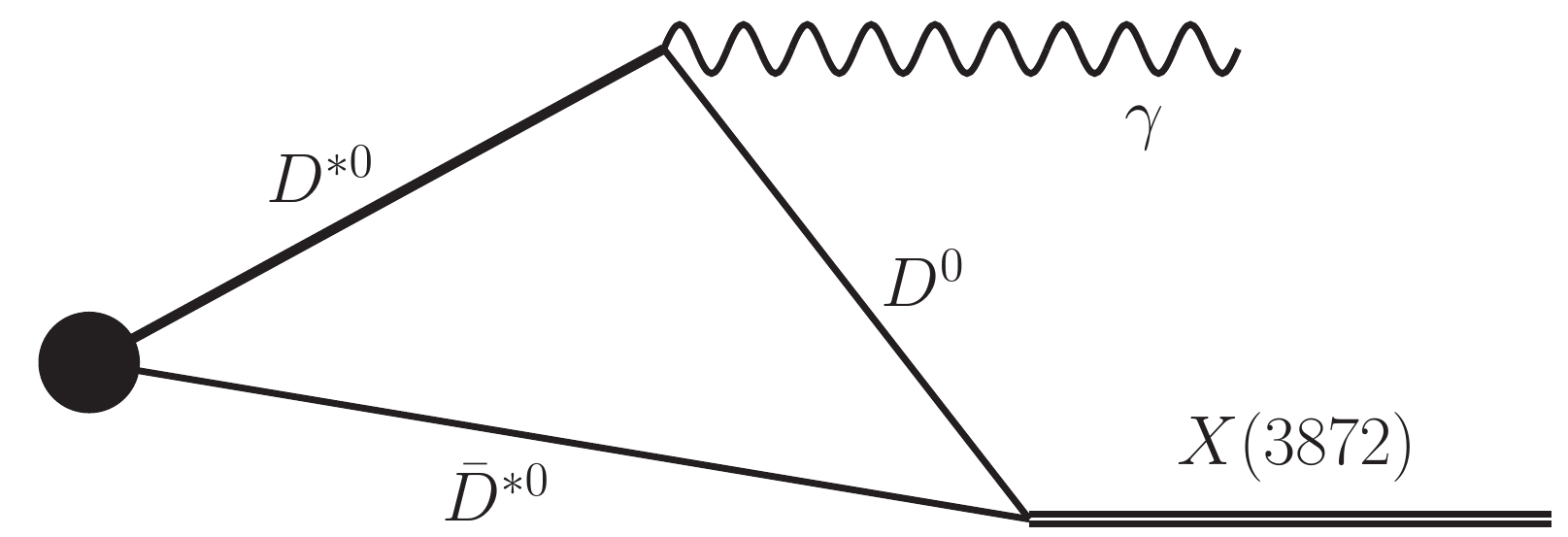}
  \caption{The mechanism for producing $\gamma \X$ from $D^{*0}\bar D^{*0}$ through a triangle diagram. The filled circle represents a short-distance source for the $D^{*0}\bar D^{*0}$ pair.}
	\label{fig:feynman}
\end{figure}

{\it Triangle singularity.}---The starting point is to produce the $\X$ and a photon from a short-distance $\dstardstar$ source produced in some reactions (here short distance means that the energy scale for producing the $\dstardstar$ pair is much larger than the one in the subsequent reaction producing $\X\gamma$, so that the $\dstardstar$ pair can be treated as a pointlike source), as shown in Fig.~\ref{fig:feynman}. 
The $S$-wave $D^{*0}\bar D^{*0}$ pair can have quantum numbers $J^{PC}=(0^{++},1^{+-},2^{++})$, and one possibility is from the $Z_c(4020)$ decays as studied in Ref.~\cite{Voloshin:2019ivc}. The key observation in this Letter is to make use of the strong sensitivity of the $\X\gamma$ line shape to the $\X$ mass around the $\dstardstar$ threshold to determine the value of $\delta$. The sensitivity is caused by a triangle singularity close to the physical region, which will be discussed below.

Triangle singularity is the leading Landau singularity~\cite{Landau:1959fi} for a triangle diagram. It is in the physical region if all the intermediate particles can go on shell and all the interaction vertices satisfy the energy-momentum conservation, known as the Coleman-Norton theorem~\cite{Coleman:1965xm}. 
Being a logarithmic branch point, the triangle singularity would produce an infrared divergence
in the reaction rate were it really in the physical region. This does never happen because at least one of the three particles must be unstable as a consequence of being on shell (so that it can decay into another intermediate particle and some external particles). As a result, the finite widths of the intermediate particles move the singularity into the complex energy plane, giving rise to a finite peak. The shape of the peak depends crucially on the location of the singularity, which is sensitive to the masses of the involved particles. 
For more details about the triangle singularity, we refer to Refs.~\cite{Eden:1966dnq,Gribov:2009zz,Anisovich:2013gha,Aitchison:2015jxa}. There have been lots of discussions about the triangle singularity in hadronic reactions since the 1960s. Its relevance to some of the recently reported resonant structures have been briefly reviewed in Refs.~\cite{Guo:2017jvc,Guo:2017wzr}.

For the $\dstardstar$ pair in an $S$ wave, because the $\X$ couples to the $\ddstaro$ also in an $S$ wave, the production amplitude of the $\X\gamma$ for the mechanism in Fig.~\ref{fig:feynman} is proportional to the scalar three-point loop integral. For the $\X\gamma$ invariant mass to be in the vicinity of the $\dstardstar$ threshold ({\it i.e.}, between about 4.01 and 4.02~GeV), the typical velocity of all the involved intermediate charmed mesons is smaller than 0.01 in the $\X\gamma$ center-of-mass frame. 
Thus, one can use nonrelativistic effective field theory (for a review, see Ref.~\cite{Guo:2017jvc}) to study this problem with tiny theoretical uncertainties. The nonrelativistic form of the scalar loop integral in the $\X\gamma$ center-of-mass frame is given by~\cite{Guo:2010ak} 
\begin{eqnarray}
    I(E_{X\gamma}) &=&   \frac{1}{E_\gamma}  
    \left[ \arctan\left(\frac{c_2-c_1}{2b\sqrt{c_1}}\right) \right. \nonumber\\
   && \left. +
\arctan\left(\frac{c_1-c_2+2b^2}{2b\sqrt{c_2-b^2}}\right) \right],
    \label{eq:Iexp}
\end{eqnarray}
where $E_\gamma=(E_{X\gamma}^2-m_X^2)/(2E_{X\gamma})$ is the photon energy, $E_{X\gamma}$ is the invariant mass of the $\X\gamma$ system, $b=m_*E_\gamma/(m_0+m_*)$, $c_1=m_*(2m_*-E_{X\gamma}-i\Gamma_*)$, and $c_2=\mu_{*0} \left[2(m_*+m_0+E_\gamma-E_{X\gamma}) +E_\gamma^2/m_0 - i\Gamma_* \right]$. Here, $m_X$, $m_*$ and $m_0$ are the masses of the $\X$, the $D^{*0}$ and the $D^0$, respectively, $\mu_{*0}=m_*m_0/(m_*+m_0)$ is the reduced mass of $D^0$ and $D^{*0}$, and $\Gamma_*$ is the $D^{*0}$ width (here it is sufficient to use a constant width~\cite{Hanhart:2010wh}).
The function has been adapted to the question at hand from the expression worked out in Ref.~\cite{Guo:2010ak}. An overall multiplicative constant factor has been dropped since it does not have any impact on the line shape. 

This function has logarithmic singularities at the solutions of
\begin{equation}
  (c_2-c_1)^2+4b^2 c_1 = 0 \, ,
  \label{eq:nrtrising}
\end{equation}
which is the nonrelativistic version~\cite{Guo:2014qra} of the Landau equation~\cite{Landau:1959fi} for the triangle singularity. 
Among all the solutions, only one of them can be close to the physical region, and we refer to Ref.~\cite{Bayar:2016ftu} for further discussions.
Applying to the case here, one finds that the relevant triangle singularity is located at
\begin{equation}
   E_{X\gamma}^\text{TS} = 2 m_* + \frac{x^2}{4m_0} + \mathcal{O}\left(\frac{x^3}{m_0^2}\right),
    \label{eq:ts}
\end{equation}
where $x=m_*-m_0-2\sqrt{-m_0\delta}+\delta$. 
Because the $D^0$ decays only weakly, and the $D^{*0}$ has a tiny decay width (given below), the triangle singularity is located in the vicinity of the physical region, providing an enhancement to the production rate, and has a dramatic impact on the $\X\gamma$ line shape.

\begin{figure}[tb]
   \includegraphics[width=\linewidth]{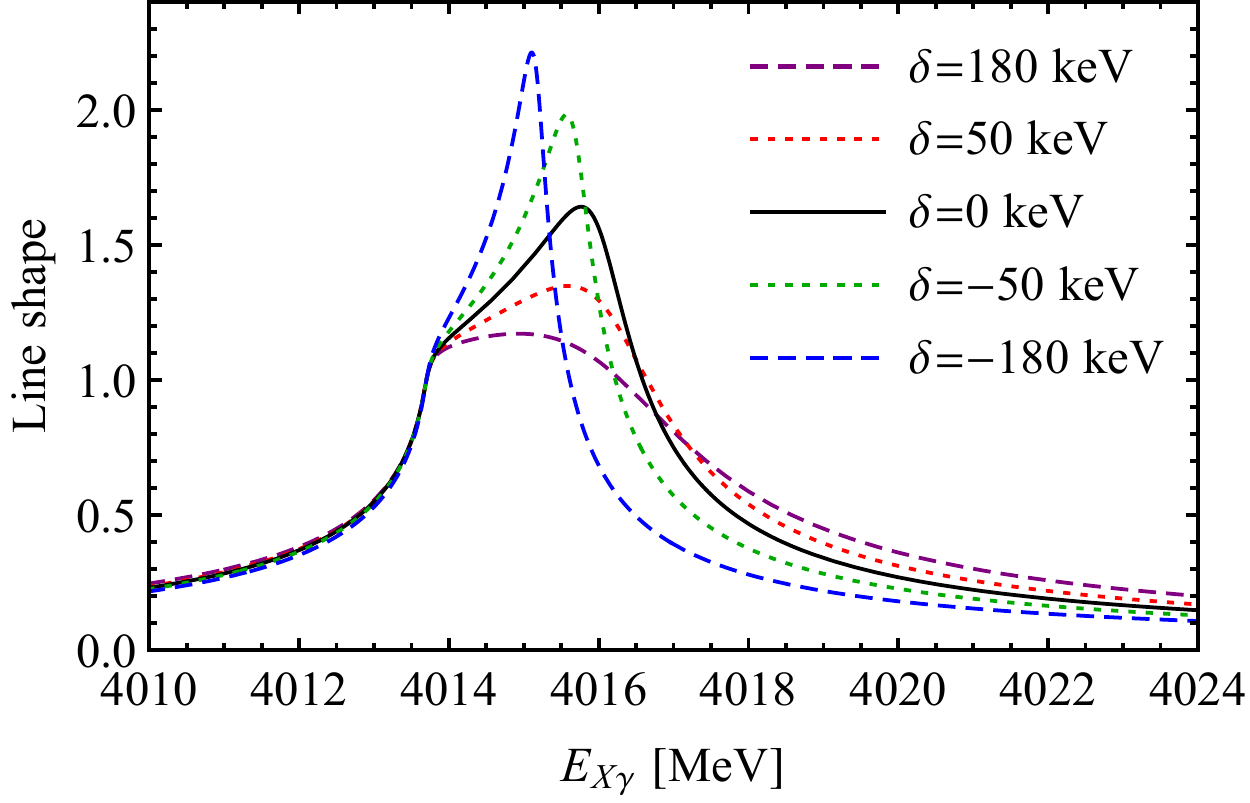}
  \caption{The $X(3872)\gamma$ line shapes in the region around the $D^{*0}\bar D^{*0}$ threshold. The solid black curve corresponds to $\delta=0$~keV; the purple and blue dashed curves correspond to $\delta=180$ and $-180$~keV, respectively; the red and green dotted curves correspond to $\delta=50$ and $-50$~keV, respectively.}
	\label{fig:ls}
\end{figure}

So far there is only an upper limit for the $D^{*0}$ width, $<2.1$~MeV~\cite{Tanabashi:2018oca}. However, the width of the charged $D^{*\pm}$ is known to a percent level, $\Gamma_{D^{*\pm}}=83.4\pm1.8$~keV~\cite{Tanabashi:2018oca}, and isospin symmetry allows us to relate the partial widths of the hadronic decays $D^{*0}\to D^0\pi^0$ and $D^{*\pm}\to D^0\pi^\pm$. Using further the branching fractions of these two decays, which are $(64.7\pm0.9)\%$ and $(67.7\pm0.5)\%$, we get the $D^{*0}$ width,
\begin{equation}
    \Gamma_* = 55.3\pm1.4~\text{keV} \,, \label{eq:gamma_dstar0}
\end{equation}
which agrees almost exactly with the result in Ref.~\cite{Rosner:2013sha}.
The line shape for the $\X\gamma$ produced in this mechanism is given by
\begin{equation}
    F(E_{X\gamma}) = \frac{|I(E_{X\gamma})|^2}{|I(2m_*)|^2} \frac{E_\gamma^3}{\left[(4 m_*^2-m_X^2)/(4m_*)\right]^3} ,
    \label{eq:ls}
\end{equation}
which has been normalized to unity at the $D^{*0}\bar D^{*0}$ threshold.
In order to clearly show the sensitivity of the line shape to the mass of the $\X$, in Fig.~\ref{fig:ls} we plot the normalized line shape for five values of $\delta$ ranging from $-180$~keV to $180$~keV. The central values of the $D^{*0}$, $D^0$ masses and $\Gamma_*$ in Eq.~\eqref{eq:gamma_dstar0} are used.  

At the $D^{*0}\bar D^{*0}$ threshold, a cusp is evident which is slightly smeared by the $D^{*0}$ width. Below the cusp, different curves almost overlap with each another. However, they behave distinctly above the threshold, as can be clearly seen from Fig.~\ref{fig:ls}. The reason is as follows. Neglecting the $D^{*0}$ width, from Eq.~\eqref{eq:ts}, one finds that for positive $\delta$ ($m_X< m_0+m_*$) the singularity is in the complex plane, while for $\delta\leq0$~keV ($m_X\geq m_0+m_*$) the singularity moves to the physical region, producing a logarithmic divergence. The finite $D^{*0}$ width moves the singularity in all cases to the complex plane, which is however still close to the physical region. The peaking energy in each curve corresponds to the real part of the singularity position, and the peak is sharper for a smaller imaginary part. For instance, using Eq.~\eqref{eq:ts} one is ready to find that for $\delta=-180$~keV the singularity is at $(4015.2-i\,0.1)$~MeV, for $\delta=-50$~keV the singularity is at $(4015.7-i\,0.2)$~MeV, and for $\delta=0$~keV the singularity moves to $(4016.0-i\,0.4)$~MeV. These singularities are responsible for the peaks above the $\dstardstar$ threshold.

{\it Sensitivity study.}---The sensitivity of the line shape on the $\X$ mass offers a portal to measure the $\X$ mass indirectly. In order to make a more quantitative statement of the sensitivity, we make a simple Monte Carlo (MC) simulation. 
First, we generate synthetic data by employing the von Neumann rejection method to select the random data points that follow the normalized distribution of Eq.~\eqref{eq:ls} for $E_{X\gamma}$ in the range from 4010 to 4020~MeV. The MC data are generated using the central values of the $D^{*0}$, $D^0$ masses and $\Gamma_*$ for three different values of $\delta$: $-50$~keV, 0~keV and 50~keV.
For each data set, the data can  then be divided into a number of bins with the statistical errors given by the square root of the number of events in each bin. The value of $\delta$ can be extracted from fitting to these synthetic data using Eq.~\eqref{eq:ls}.
Varying the MC event numbers and the bin widths, one can investigate the impact of the event number as well as the bin width on the precision of the so-extracted $\X$ mass. 

\begin{table*}[tb]
  \caption{Results on $\delta$ (in units of keV) from fitting to the synthetic data sets generated using three different input values of $\delta_\text{in}$ : $-50$~keV, 0~keV, and $50$~keV.  \label{tab:fitresult} }	
		\begin{ruledtabular}
		\begin{tabular}{l c c c  }
		 Data set $A$ &  $\delta_\text{in}=-50$ keV (127 events) & $\delta_\text{in}=0$ keV (164 events) & $\delta_\text{in}=50$ keV (192 events) \\\hline
		 10 bins & $-24^{+24}_{-28}$ & $11^{+31}_{-20}$ & $22^{+41}_{-23}$ \\ 
		 5 bins & $-17^{+24}_{-27}$ & $30^{+64}_{-29}$ & $40^{+67}_{-31}$ \\
		 \hline\hline
		 Data set $B$ & $\delta_\text{in}=-50$ keV (626 events) & $\delta_\text{in}=0$ keV (831 events) & $\delta_\text{in}=50$ keV (1006 events) \\\hline
		 10 bins & $-47^{+13}_{-16}$ & $-1^{+13}_{-11}$ & $63^{+34}_{-24}$ \\
		 5 bins &  $-48^{+15}_{-19}$ & $-4^{+11}_{-10}$ & $53^{+38}_{-25}$ \\
         \hline\hline
          Data set $C$ & $\delta_\text{in}=-50$ keV (3133 events) & $\delta_\text{in}=0$ keV (4027 events) & $\delta_\text{in}=50$ keV (5015 events) \\\hline
		 10 bins &  $-53^{+7}_{-8}$ & $-2\pm5$ & $55^{+13}_{-11}$ \\
		 5 bins &  $-52^{+7}_{-8}$ & $-2^{+7}_{-6}$ & $61^{+17}_{-14}$ \\
		\end{tabular}
		\end{ruledtabular}
\end{table*}

\begin{figure*}[bt]
   \includegraphics[width=0.32\textwidth]{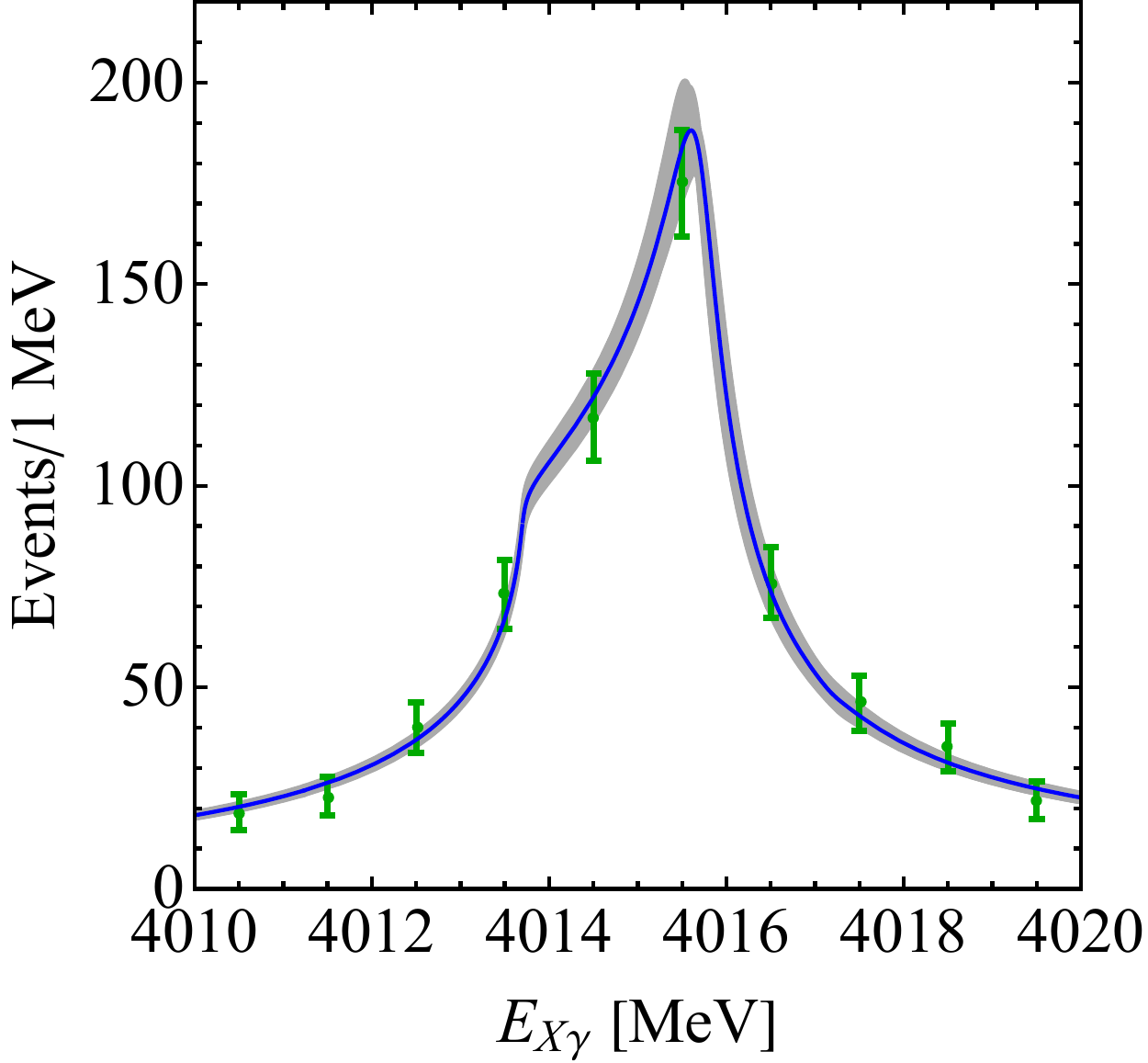}\hfill
   \includegraphics[width=0.32\textwidth]{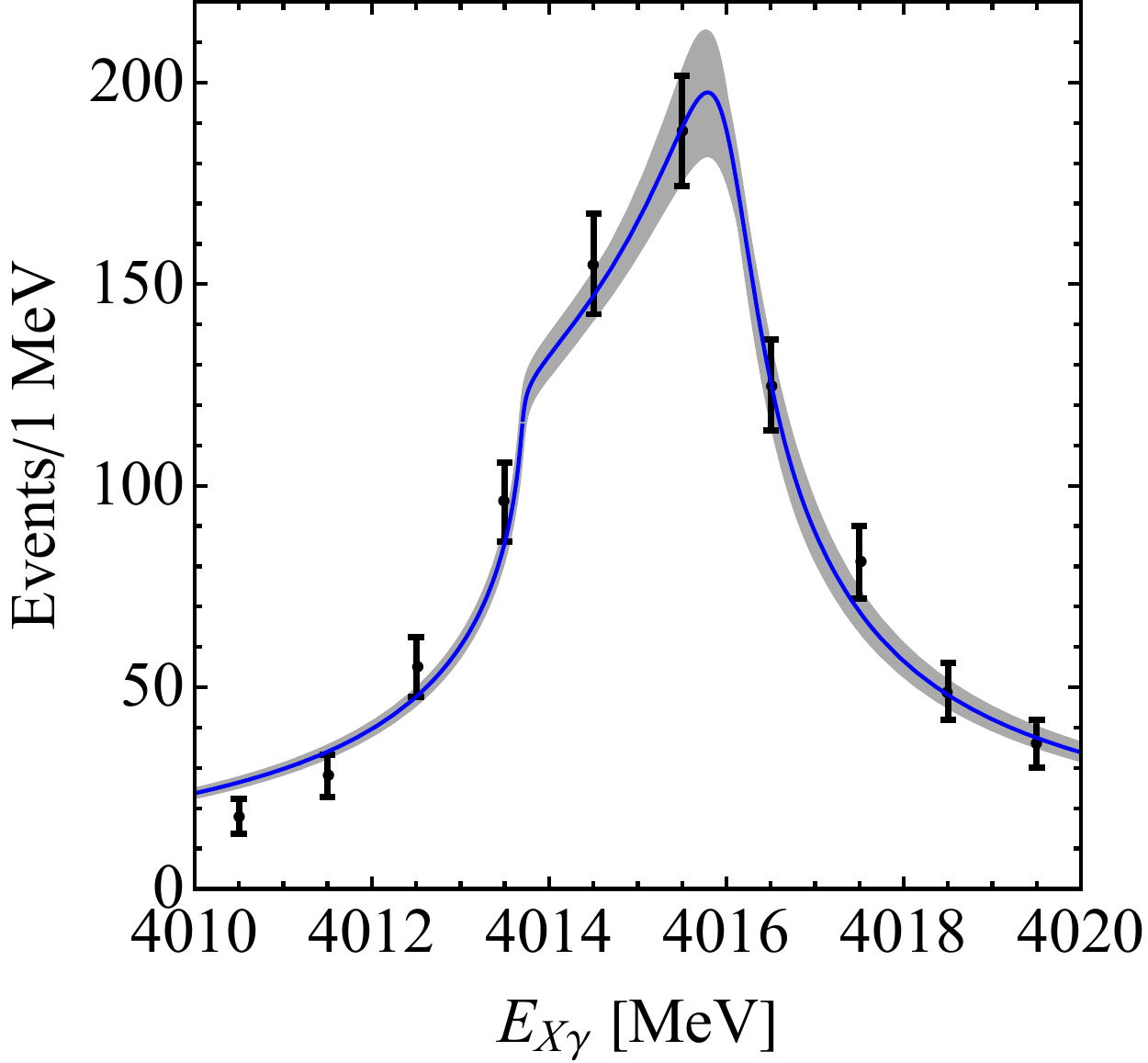}\hfill
   \includegraphics[width=0.32\textwidth]{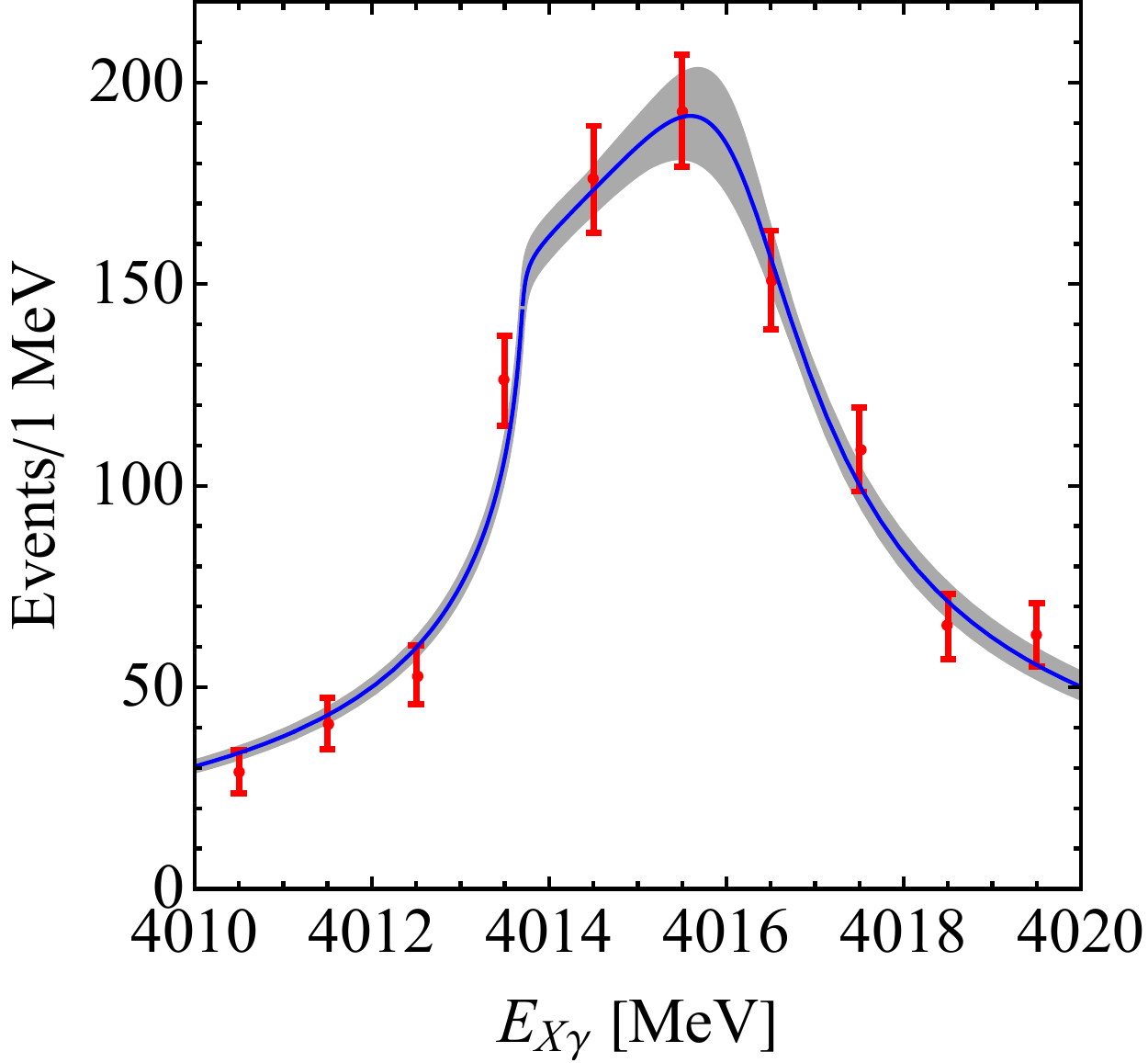}\\[5mm]
   \includegraphics[width=0.32\textwidth]{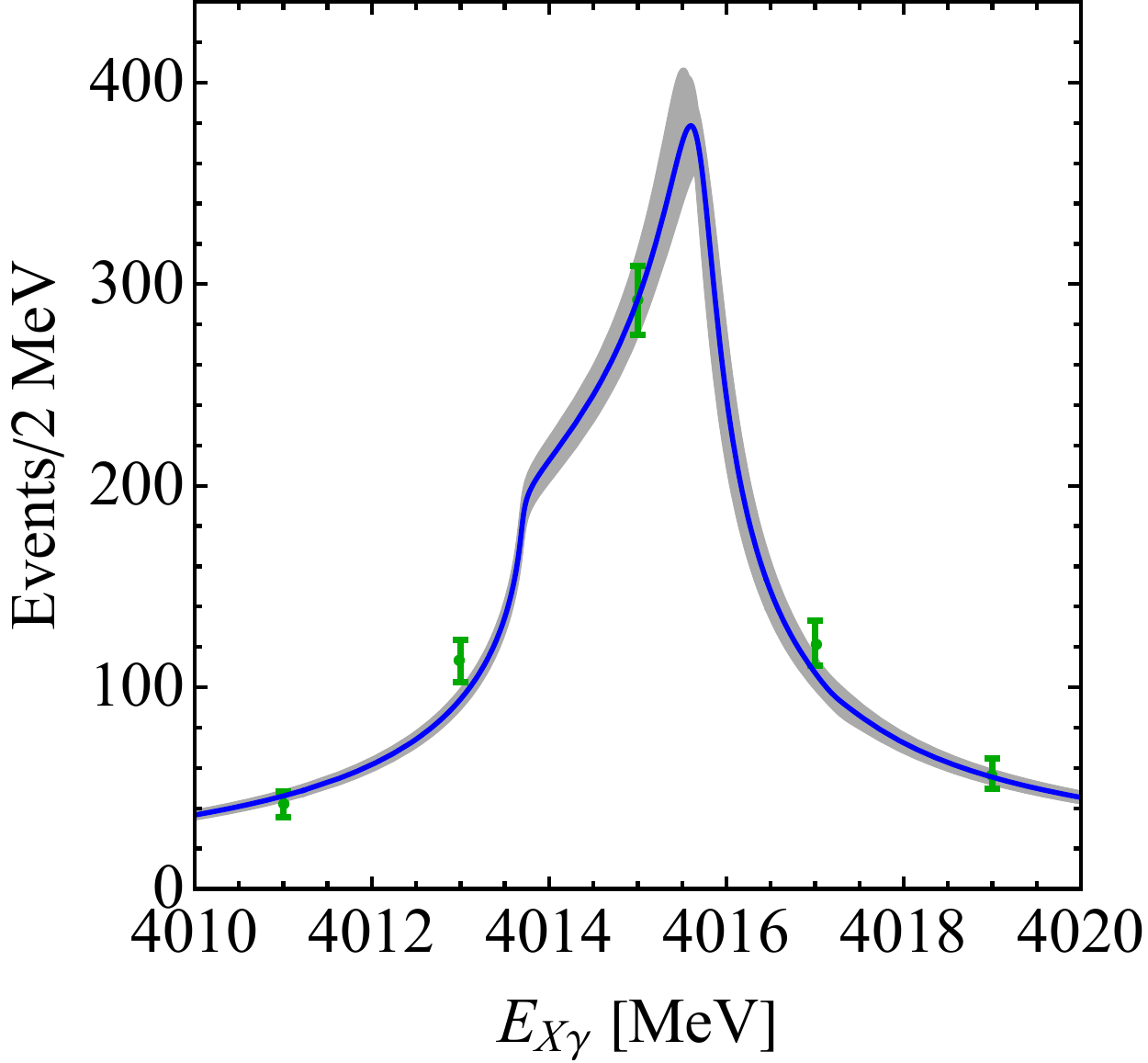}\hfill
   \includegraphics[width=0.32\textwidth]{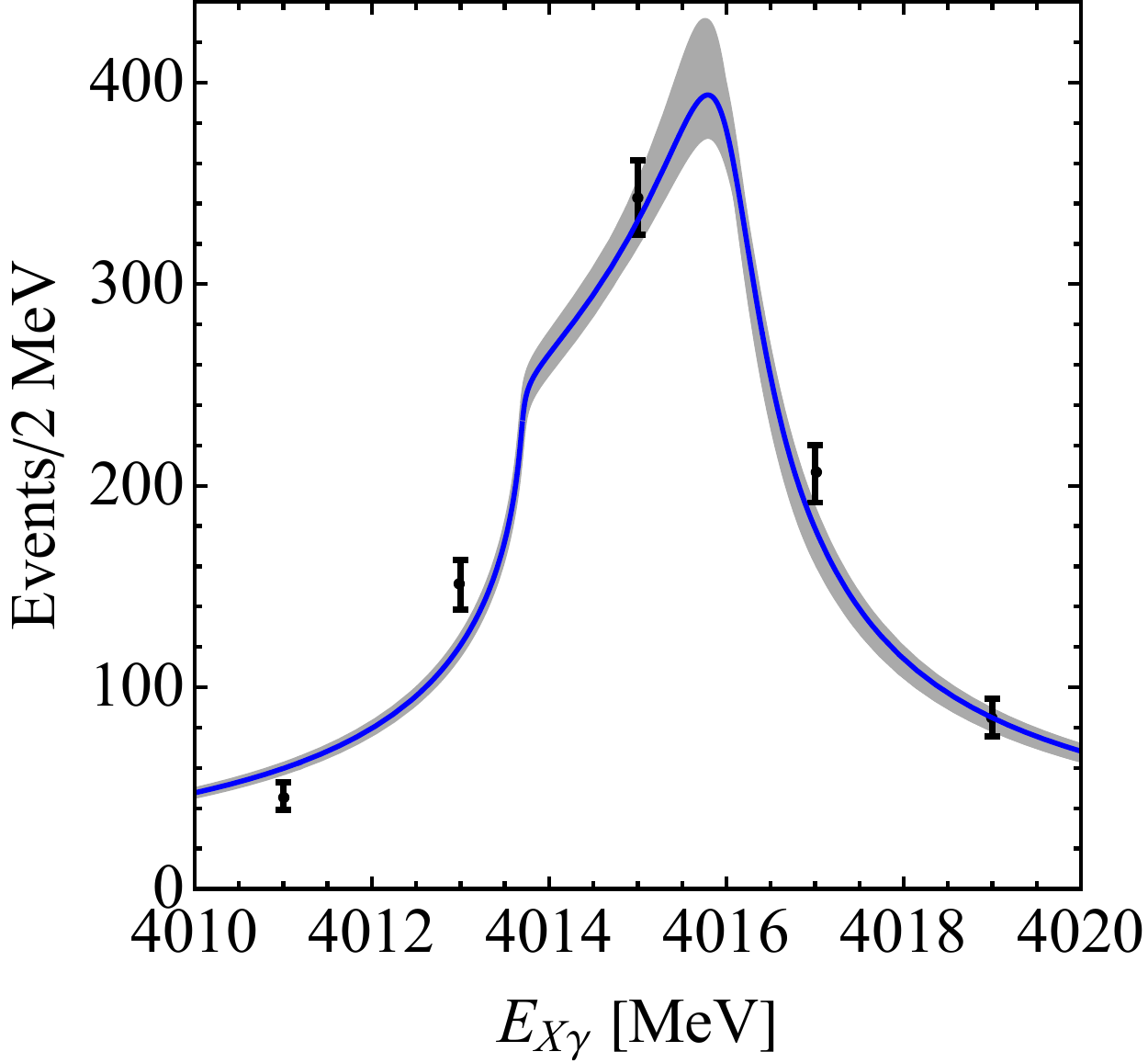}\hfill
   \includegraphics[width=0.32\textwidth]{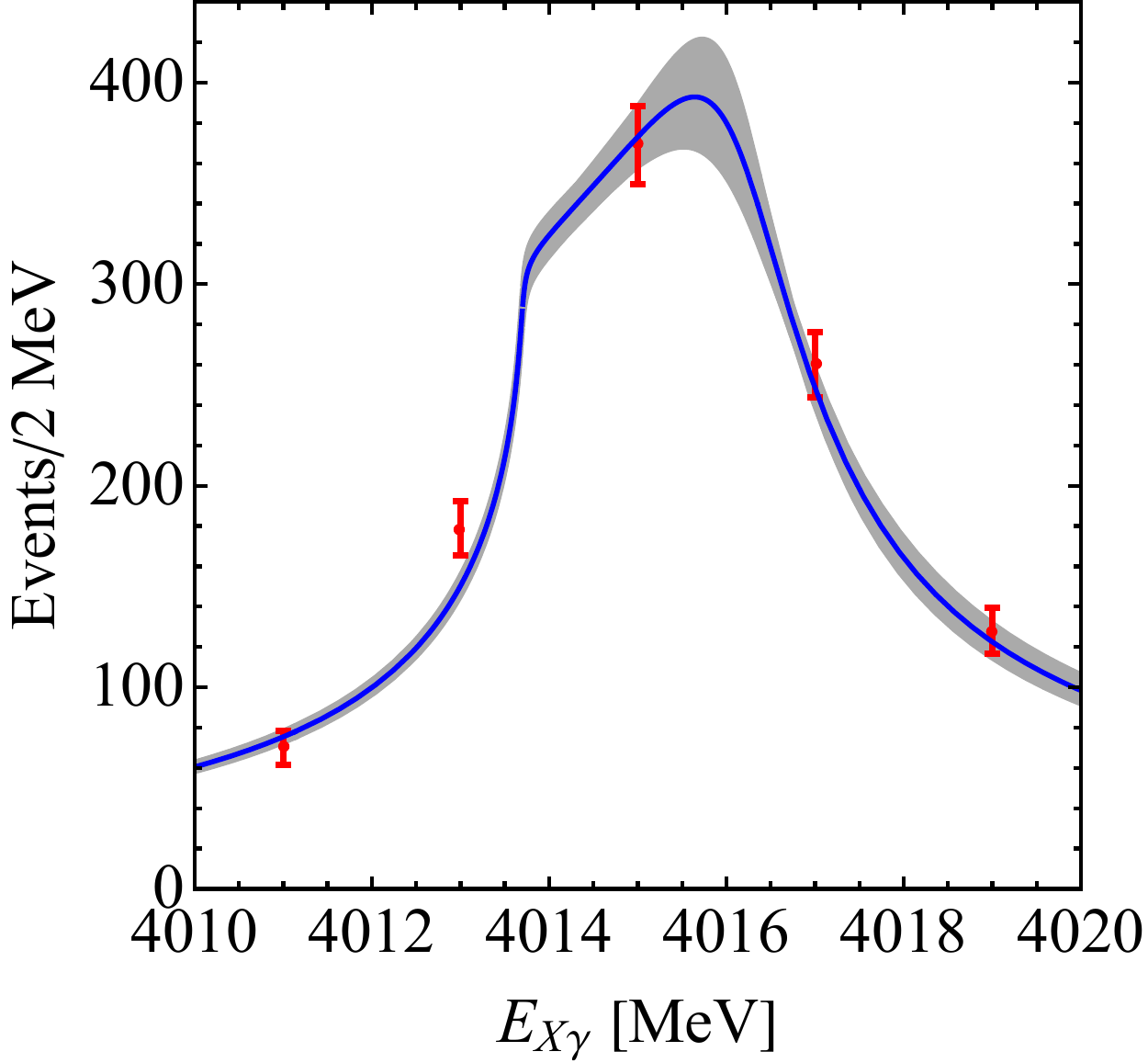}
  \caption{The synthetic data set $B$ generated using $\delta_\text{in}=-50$~keV (left), $\delta_\text{in}=0$~keV (middle), and $\delta_\text{in}=50$~keV (right) and the corresponding best fits. In the upper and lower panels, the events are divided into ten and five bins, respectively. The bands are the $1\sigma$ uncertainty region from fitting.}
	\label{fig:fits}
\end{figure*}

As examples, we generated different data sets ($A$, $B$, and $C$) using  the input value $\delta_\text{in}=-50$~keV, 0~keV and 50~keV. The event numbers in data set $C$ are about 5 times those in data set $B$, which are again about 5 times those in data set $A$. The event numbers in each data set are given in Table~\ref{tab:fitresult}. The synthetic events in each data set for each $\delta_\text{in}$ are divided into ten bins and five bins, respectively. The fitting results for data set $B$, together with the synthetic data, are shown in Fig.~\ref{fig:fits}, and the extracted $\delta$ values from all data sets are collected in Table~\ref{tab:fitresult}. 
The first observation is that with a few hundred events in this energy region, one can achieve a precision much higher than the current one ($\pm180$~keV). With about 600 events for the input $\delta=-50$~keV and about 1000 events for $\delta=50$~keV, one is ready to know the sign of $\delta$, i.e., whether the $\X$ is above or below the $\ddstaro$ threshold,  with a confidence level larger than 95\%. Another important point is that the precision changes only marginally with a bin size of 2~MeV compared with that with a bin size of 1~MeV, in particular for negative $\delta_\text{in}$ which produces a more pronounced peak. The reason is that the locations of both the $\dstardstar$ threshold cusp and the triangle singularity are fixed once all the involved masses are known, and the line shape is largely determined by them. Such a weak sensitivity to bin sizes in the case of using the threshold cusp to extract the $\pi\pi$ scattering length has been noticed in Ref.~\cite{Liu:2012dv}. It is worthwhile to mention that the precision does not depend on the uncertainties of the $D^{*0}$ and $D^0$ masses.
This is because it is the mass difference $\delta$ that enters into the game.

{\it Conclusion.}---In this Letter, a novel method to measure the $\X$ mass precisely is proposed. This is an indirect method in the sense that it does not involve measuring the line shape of the $\X$, but involves measuring the line shape of the $\X\gamma$ pair. This method requires the $\X\gamma$ to be produced from a short-distance $\dstardstar$ source.
The $S$-wave $\dstardstar$ pair can have quantum numbers of $J^{PC}=(0^{++},1^{+-},2^{++})$. The $\X$ needs be reconstructed in modes other than the $D^0\bar D^0\pi^0$, which include the $J/\psi\pi^+\pi^-$, $J/\psi\pi^+\pi^-\pi^0$, $J/\psi\gamma$ and $\psi'\gamma$. Otherwise, one needs to take into account the interference between the triangle diagram with the contribution of $D^{*0}\bar D^{*0}\to\gamma D^0\bar D^0\pi^0$ without the $\X$ in the intermediate state.
The method can be applied at electron-position facilities such as BES-III and possible future super charm-tau facilities, where the $\X\gamma$ pair needs to be produced associated with another positive $C$-parity neutral meson, e.g., $e^+e^-\to \dstardstar\pi^0\to \pi^0\X\gamma$. This can also be applied to proton-antiproton reactions at $\overline{\text{P}}$ANDA at energies around the $\dstardstar$ threshold, which is particularly promising. If the event number in an exclusive reaction is not enough, one may even try to measure the $\X\gamma$ line shape inclusively, as other mechanisms provide only a smooth background without a nearby singularity. Even if the experimental events contain another nontrivial structure in a nearby energy region, such as the $Z_c(4020)$, the formalism can be adapted to take that into account. 

At last, let me emphasize again that the proposed method measures directly the difference between the $\X$ mass and the $\ddstaro$ threshold, and the measurement depends little on the uncertainties of the $D^{*0}$ and $D^0$ masses, which are 50~keV for both. Therefore, it can lead to a much more precise measurement of $\delta$ than the usual method of measuring the $\X$ line shape. An improved measurement of the $\X$ mass is foreseen, and it will be an important step towards a deeper understanding of the nature of the $\X$, and eventually of other related $XYZ$ states.

\vspace{0.5cm}

\begin{acknowledgements}
I would like to thank Simon Eidelman for inviting me to give a talk at the PhiPsi2019 Workshop, which provides a lively discussion environment and during which this work is done. I am grateful to Christoph Hanhart and Cheng-Ping Shen for useful comments.
This Letter is supported in part by the National Natural Science Foundation of China (NSFC) and  the Deutsche Forschungsgemeinschaft (DFG) through the funds provided to the Sino-German Collaborative Research Center ``Symmetries and the Emergence of Structure in QCD"  (NSFC Grant No. 11621131001, DFG Grant No. TRR110), by the NSFC under Grants No. 11747601 and 11835015, by the Chinese Academy of Sciences (CAS) under Grants No. QYZDB-SSW-SYS013 and XDPB09, and by
the CAS Center for Excellence in Particle Physics (CCEPP).
\end{acknowledgements}

\bibliography{x}

\end{document}